\documentclass{article}

\usepackage{amsmath,amssymb}

\usepackage{graphicx}

\def\TT{{\mathbb T}}
\def\ZZ{{\mathbb Z}}
\def\RR{{\mathbb R}}
\def\CC{{\mathbb C}}

\def\scrN{{\mathcal N}}

\def\i{\textrm{i}}

\def\lcm{\operatorname{lcm}}

\def\Tr{\operatorname{Tr}}

\begin{document}

\title{Intermediate statistics in quantum maps}

\author{Olivier Giraud$^1$, Jens Marklof$^2$ and Stephen O'Keefe$^2$\\[3mm]
$^1$ H.H.Wills Physics Laboratory, University of Bristol\\ 
Tyndall Avenue, Bristol BS8 1TL, U.K.\\
$^2$ School of Mathematics, University of Bristol\\
University Walk, Bristol BS8 1TW, U.K.}

\maketitle

\begin{abstract}
We present a one-parameter family of quantum maps 
whose spectral statistics are of the same intermediate type
as observed in polygonal quantum billiards. Our central result is the
evaluation of the spectral two-point correlation
form factor at small argument, which in turn yields the
asymptotic level compressibility for macroscopic correlation lengths. 
\end{abstract}



\section{Introduction}

The classification of quantum systems according to universal statistical
properties is one of the central objectives in the study of quantum chaos.
It is generally believed that the spectral statistics
of systems with chaotic classical limit are governed by
random matrix ensembles, while systems with integrable 
classical dynamics follow the statistical properties of
independent random variables from a Poisson process 
\cite{BerTab77,BohGanSch84,Haa01}. 
Interestingly, certain billiards in rational polygons 
fall in neither of the two universality classes:
the energy level correlations are conjectured to be
of intermediate type 
\cite{BogGerSch99,CasPro99,BogGerSch01,BogGirSch01,GorWie03}.
In particular, this means that
the consecutive level spacing distribution $P(s)$ exhibits level repulsion 
similar to random matrix eigenvalues, but has an exponential tail 
as for independent random variables. Furthermore, 
the spectral form factor $K_2(\tau)$ is intermediate between
0 and 1 in the limit $\tau\to 0$. One standard example 
for a statistics of this type is the semi-Poisson distribution, for
which $P(s)=4s\exp(-2s)$ and $K_2(\tau\to 0)=1/2$.

\begin{figure}
\begin{center}
\includegraphics[width=0.33\textwidth,angle=-90]{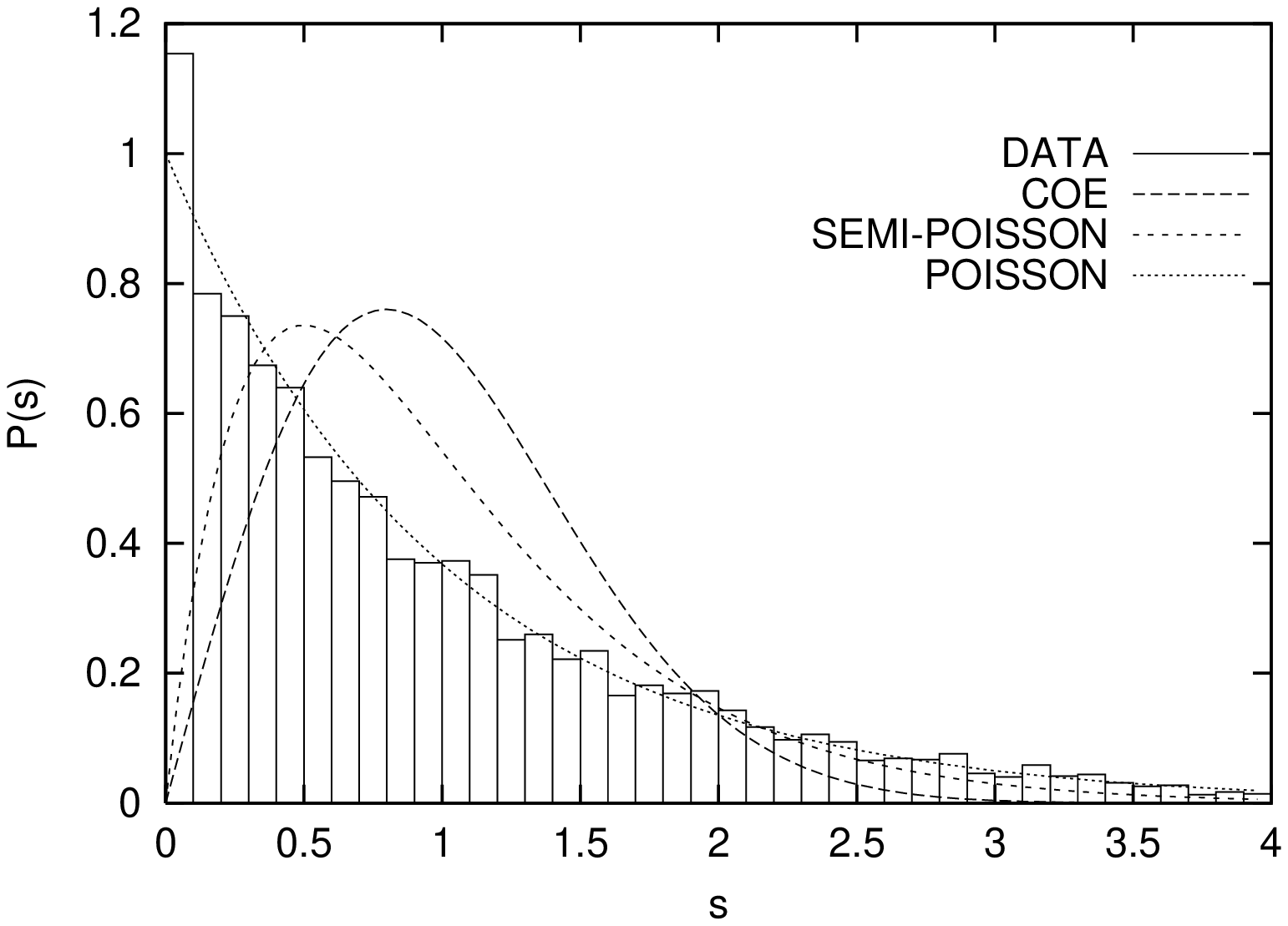}
\includegraphics[width=0.33\textwidth,angle=-90]{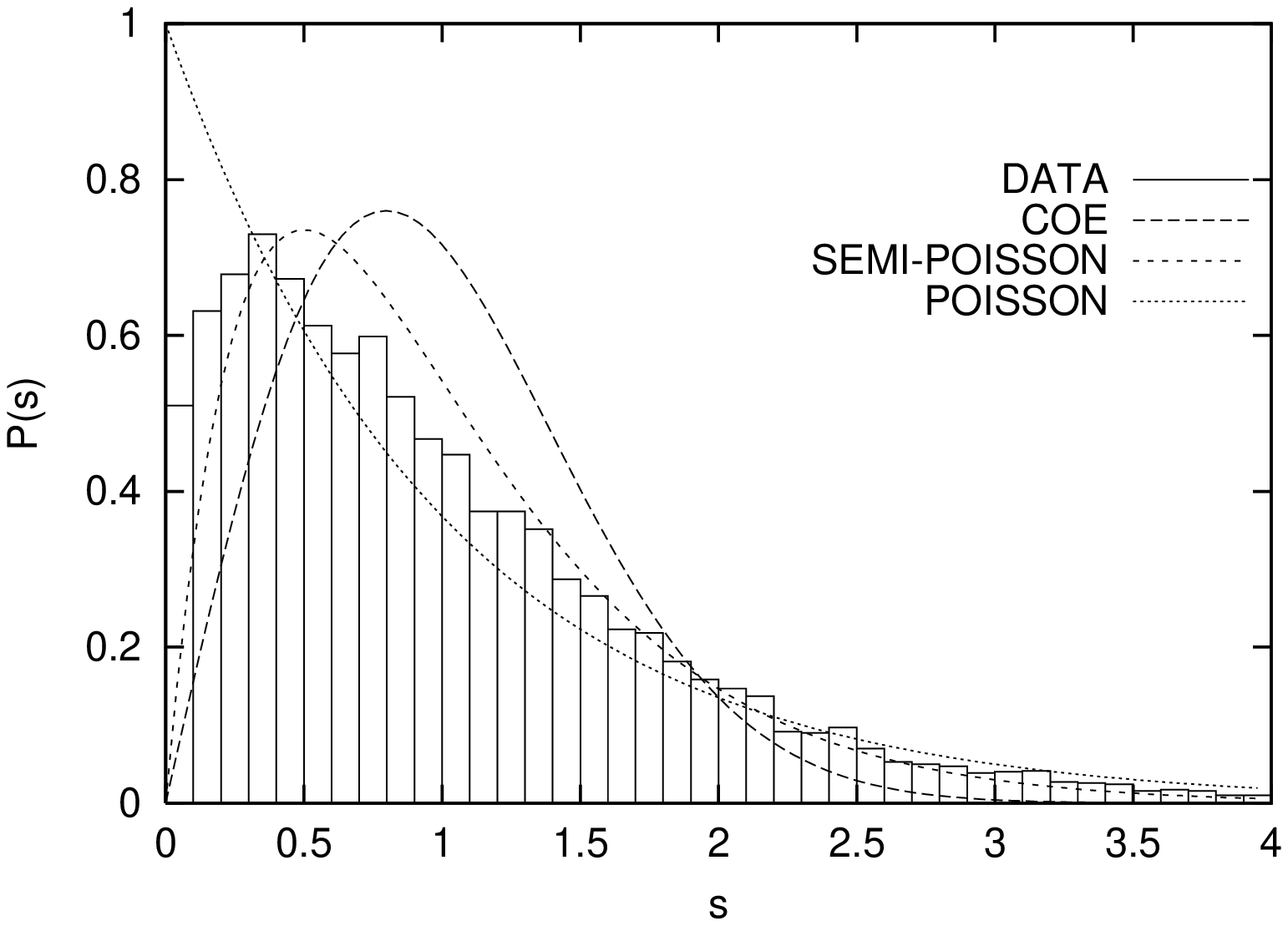}
\caption{\label{Ps1} The consecutive level spacing distribution
for Hilbert space dimension $N=7001$ and
$\alpha=1/2$ (left) and $\alpha=2/3$ (right).
The Poisson distribution corresponds to $P(s)=\exp(-s)$,
semi-Poisson to $P(s)=4s\exp(-2s)$, and COE
to the level spacing distributions of the circular orthogonal
random matrix ensembles.}
\end{center}
\end{figure}
\begin{figure}
\begin{center}
\includegraphics[width=0.33\textwidth,angle=-90]{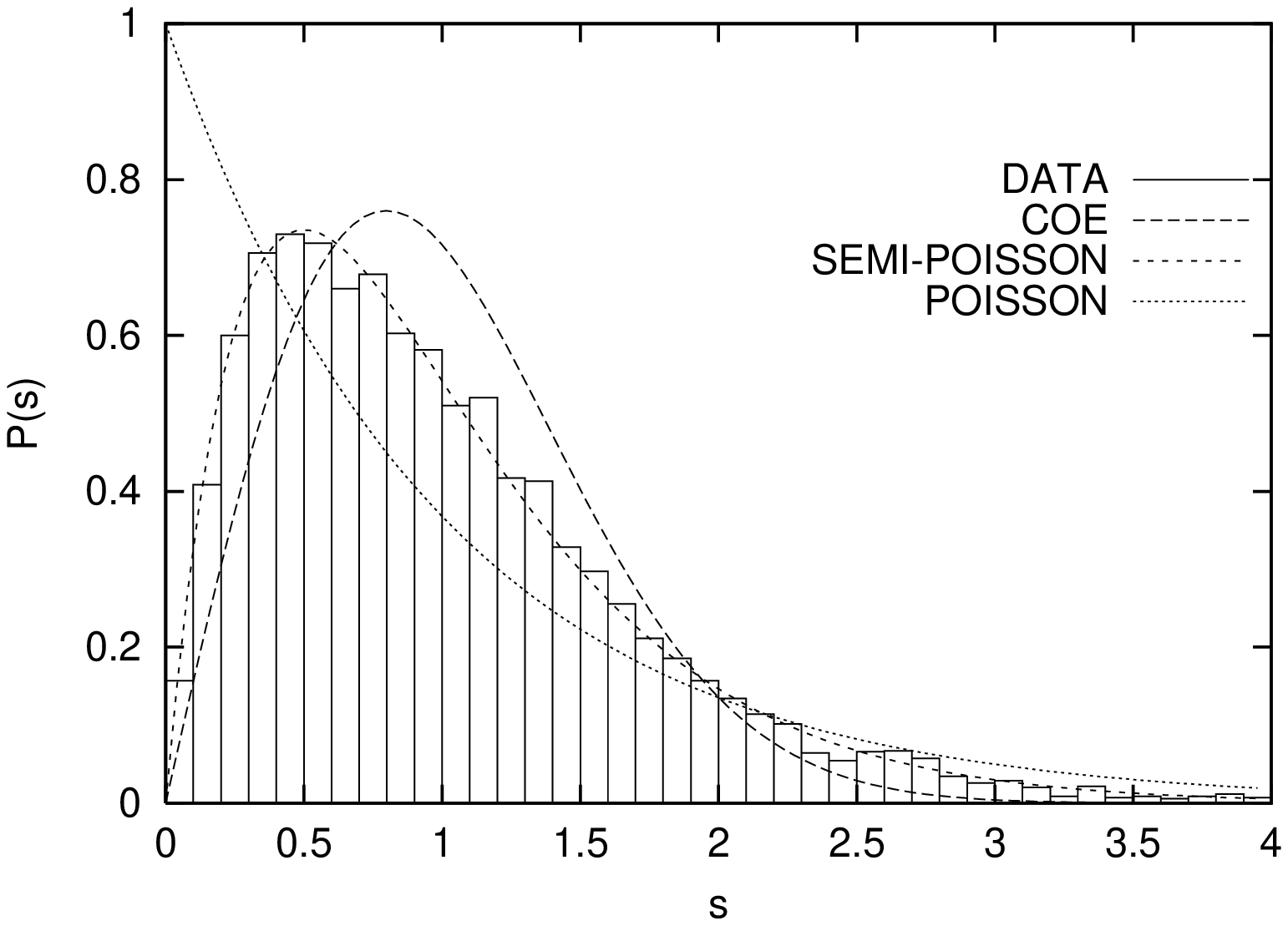}
\includegraphics[width=0.33\textwidth,angle=-90]{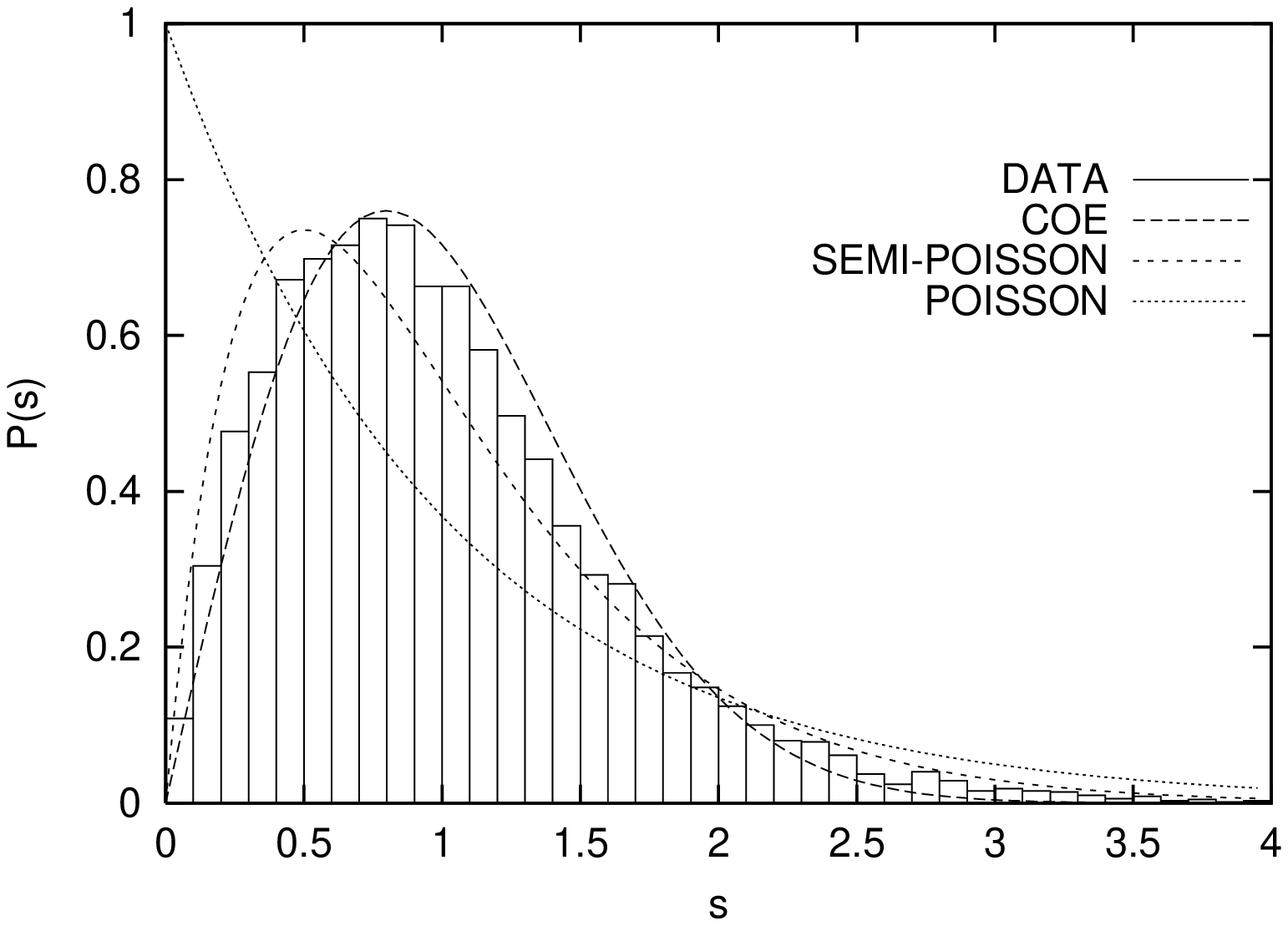}
\caption{\label{Ps2} The consecutive level spacing distribution
for Hilbert space dimension $N=7001$ and
$\alpha=3/5$ (left) and $\alpha=5/8$ (right).}
\end{center}
\end{figure}
\begin{figure}
\begin{center}
\includegraphics[width=0.33\textwidth,angle=-90]{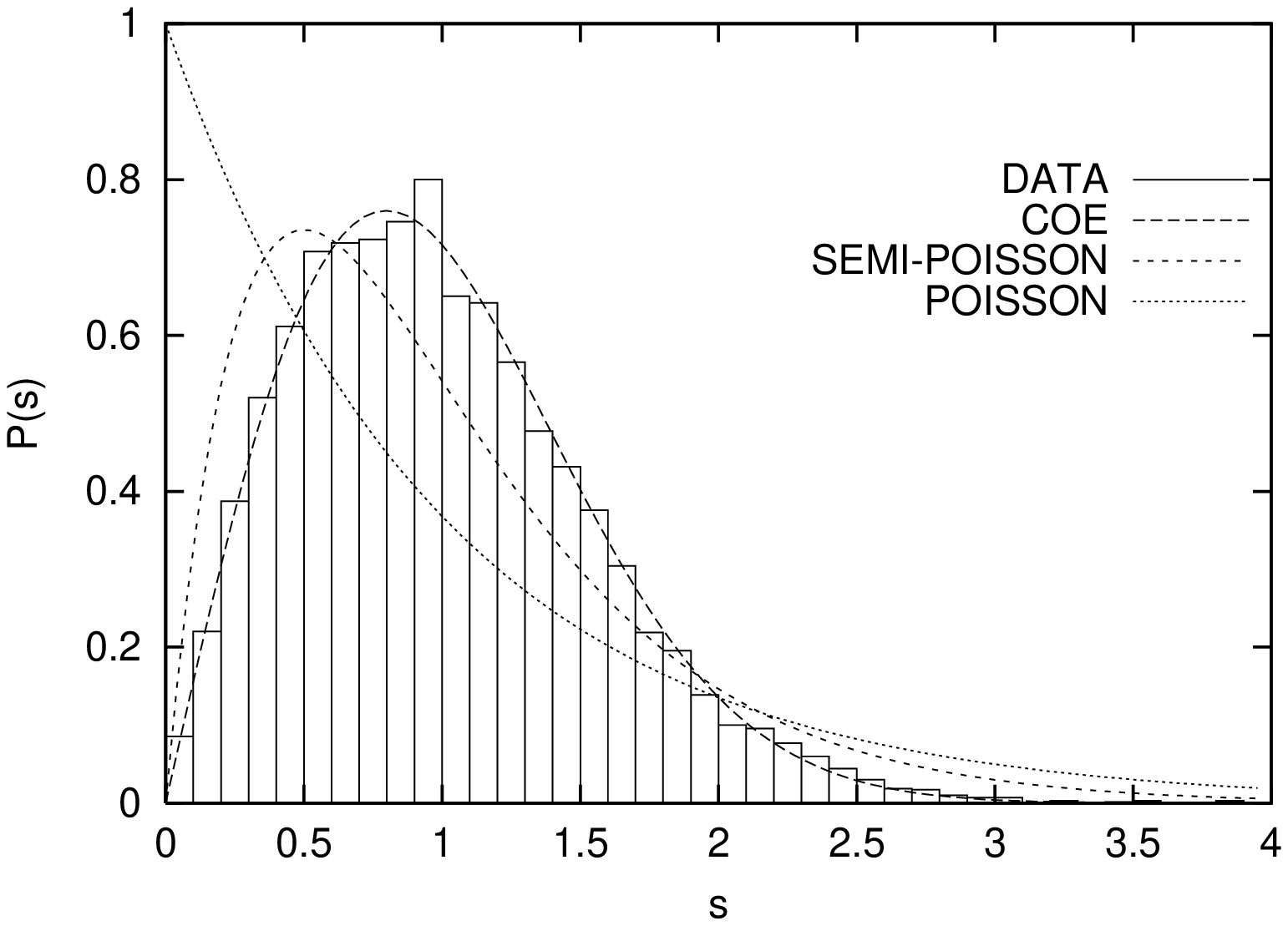}
\includegraphics[width=0.33\textwidth,angle=-90]{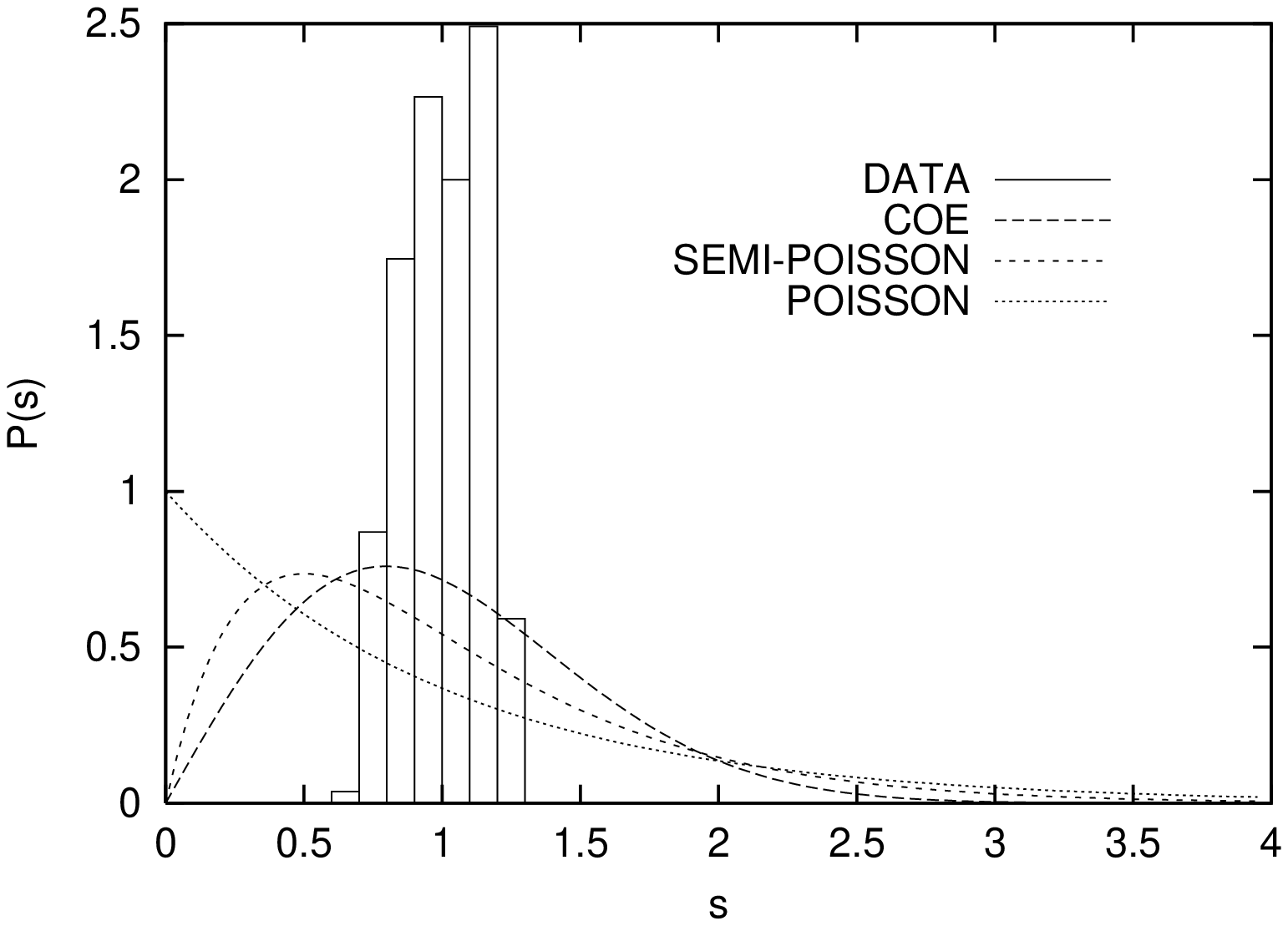}
\caption{\label{Ps3} The consecutive level spacing distribution
for $\alpha=(\sqrt{5}-1)/2$ and Hilbert space dimensions $N=6997$ (left) and $N=5867$
(right) corresponding to values of $\epsilon\sqrt N\simeq 32.1$ and 
$\epsilon\sqrt N=0.41$, respectively.}
\end{center}
\end{figure}

In this paper we present a one-parameter family of quantum maps 
whose spectral statistics are of a similar intermediate type
as observed in polygonal billiards, cf.~Figs.~\ref{Ps1} and \ref{Ps2}.  
The main result of our investigation is the evaluation
of the spectral form factor $K_2(\tau)$ at small argument. 
It is based on a number-theoretic analysis
which turns out to be considerably easier
than the geometric approach required for billiards \cite{BogGirSch01}. 

Consider the following map of the two-torus $\TT^2=\RR^2/\ZZ^2$,
\begin{equation}
\Phi_f: \TT^2 \to \TT^2, \quad \begin{pmatrix} p \\ q \end{pmatrix}
\mapsto
\begin{pmatrix}p + f(q) \\ q + 2(p+f(q)) \end{pmatrix}   
\end{equation}
where $f$ is some $1$-periodic function.
The map is a concatenation $\Phi_f=\Phi_0\circ\rho_f$ of
free motion $\Phi_0$ and kick $\rho_f$,
\begin{equation}
\Phi_0: \begin{pmatrix} p \\ q \end{pmatrix}
\mapsto
\begin{pmatrix}p  \\ q + 2p \end{pmatrix},  \quad
\rho_f : \begin{pmatrix} p \\ q \end{pmatrix}
\mapsto
\begin{pmatrix}p + f(q) \\ q \end{pmatrix} .
\end{equation} 
The quantization of a torus map
associates with it a unitary operator acting on the 
$N$-dimensional Hilbert space
of functions $\psi: \ZZ_N \to \CC$ with inner product
$\langle \psi| \phi \rangle = N^{-1} \sum_{Q=0}^{N-1} \psi^*(Q) \phi(Q)$.
Here $\ZZ_N=\ZZ/N\ZZ$ denotes the integers modulo $N$, and $N$ has the 
physical interpretation of an inverse Planck's constant.
The quantum evolution operators $U(\Phi_0)$ and $U(\rho_f)$ 
corresponding to $\Phi_0$ and $\rho_f$, respectively, are 
defined by the matrix elements 
(cf.~\cite{BerBalTabVor79,HanBer80,Isr86,BouDeB96})
\begin{equation}
\label{u0}
\langle Q'|U(\Phi_0)|Q \rangle 
= \frac{1}{N} \sum_{P=0}^{N-1} e_N\left(-P^2+P(Q'-Q) \right),
\end{equation}
\begin{equation}
\label{rf}
\langle Q'|U(\rho_f)|Q \rangle =
\langle Q'|Q \rangle \;e\left(- N V \left(\tfrac{Q}{N}\right)\right),
\end{equation}
where $V(q)$ is a periodic function defined by $f(q)=-V'(q)$, 
and $e_N(x)=\exp(2\pi\i x/N)$ and $e(x)=\exp(2\pi\i x)$.
Furthermore $U(\Phi_f)=U(\Phi_0)U(\rho_f)$ and thus
\begin{equation}
\label{uf}
\langle Q'|U(\Phi_f)|Q \rangle 
= \langle Q'|U(\Phi_0)|Q \rangle \;
e\left(- N V \left(\tfrac{Q}{N}\right)\right).
\end{equation}
We are interested in the special case of the piecewise linear 
sawtooth potential 
$V(q)=-\alpha\{q\}$ for some real constant $\alpha$,
where $\{q\}$ denotes the fractional part of $q$.
In this case $f(q)=\alpha$. The corresponding classical map 
$\Phi_\alpha:=\Phi_f$ is
uniquely ergodic for irrational $\alpha$ (in particular, there
are no periodic orbits) but not mixing. 
For rational $\alpha$, the motion can be 
identified with an interval-exchange transformation.
Note that in the momentum representation $|P \rangle$, with 
$\langle Q|P \rangle=N^{-1/2} e_N(PQ)$, the operator $U(\Phi_\alpha)$ 
has the representation
\begin{equation} \label{PUP}
\langle P'|U(\Phi_\alpha)|P \rangle
=  \frac1N \, e_N(-{P'}^2) \,\frac{1-e(N\alpha)}{1-e_N(P-P'+N\alpha)}
\end{equation}
if $N\alpha\notin\ZZ$ and 
\begin{equation}\label{PUP0}
\langle P'|U(\Phi_\alpha)|P \rangle
=e_N(-{P'}^2) \langle P' | P + N\alpha \rangle
\end{equation}
otherwise. 

Since $U(\Phi_\alpha)$ is unitary its eigenvalues are of the form 
$\exp(\i\theta_j)$ with eigenphases
$0\leq \theta_1 \leq \theta_2 \leq \cdots \leq \theta_N < 2\pi$;
it is convenient to set $\theta_0:=\theta_N-2\pi$.
The spacing distribution for consecutive levels is described by
the probability density
\begin{equation}
P(s)= \frac1N \sum_{j=1}^N 
\delta\left(s-\frac{N}{2\pi}(\theta_j-\theta_{j-1})\right),
\end{equation}
where the factor of $N/2\pi$ 
ensures that $s$ measures spacings on the scale of the
mean level spacing $2\pi/N$. 
Figs.~\ref{Ps1}-\ref{Ps3} display the spacing distribution $P(s)$
of the eigenphases of the matrix  $U(\Phi_\alpha)$  for both 
rational (Figs.~\ref{Ps1}-\ref{Ps2}) and irrational (Fig.~\ref{Ps3})
values of $\alpha$, and $N$ a prime number. 
In the case of rational $\alpha=a/b$ one should avoid 
Hilbert space dimensions $N$ divisible by $b$, since 
the matrix \eqref{PUP0} has highly singular statistics \cite{MarRud00,BacHaa99}. 
For irrational $\alpha$, we find in 
Fig.~\ref{Ps3} (left) a spacing distribution
which resembles those of random matrices from the 
the Circular Orthogonal Ensemble (COE). COE statistics are normally expected
for systems with chaotic classical limit and time-reversal symmetry
\cite{BohGanSch84,Haa01}. In our case the time-reversal transformation
which anti-commutes with $U(\Phi_\alpha)$ is 
$T'=U(\Phi_0)^{1/2}\,T\, U(\Phi_0)^{-1/2}$, where
$T$ denotes the complex conjugation operator $T\psi:=\psi^*$.
Localized spacing distributions of the type seen in Fig.~\ref{Ps3} (right)
occur when $\epsilon$, defined as
the oriented distance of $N\alpha$ to the nearest integer, is
of the order $1/\sqrt N$. Such correlations arise as the perturbation of
a rigid spectrum, and will be described in section \ref{secNon}. 

The map $\Phi_\alpha$ has in fact a further classical symmetry: it
commutes with $\rho_{1/2}$ for any $\alpha$. Thus,
for $N$ even, the corresponding operators $U(\Phi_\alpha)$ 
and $U(\rho_{1/2})$
commute and the eigenstates of $U(\Phi_\alpha)$ fall into
two parity classes according to $U(\rho_{1/2})\varphi_j=\varphi_j$
or $U(\rho_{1/2})\varphi_j=-\varphi_j$, respectively.
Our numerical experiments suggest
that the level statistics for each subspectrum for even $N$ are of the same
type as those for odd $N$ displayed in Figs.~\ref{Ps1}-\ref{Ps3}.

A statistics which is more accessible from an 
analytical point of view is the two-point 
correlation density (which describes the distribution of {\em all}
spacings)
\begin{equation}\label{R2}
R_2(s)= \frac1N \sum_{j,k=1}^N \sum_{m\in\ZZ}
\delta\left(s-\frac{N}{2\pi}(\theta_j-\theta_k+2\pi m)\right) .
\end{equation}
The Poisson summation formula applied to the $m$-sum yields
\begin{equation}\label{R22}
R_2(s)=\frac{1}{N^2}\sum_{n\in\ZZ}
\big|\Tr \left[U(\Phi_\alpha)^n\right]\big|^2 e_N(n s).
\end{equation}
The spectral form factor is defined as the Fourier transform of $R_2(s)$,
\begin{equation}
\label{formfactor}
K_2(\tau)=\frac{1}{N}\big|\Tr \left[U(\Phi_\alpha)^n\right]\big|^2,
\quad
\tau=n/N .
\end{equation}
In the following section we will calculate
\begin{equation} \label{K00}
\overline{K_2(0)}
:=
\lim_{n\to\infty} \lim_{N_\nu\to\infty}
\frac1n \sum_{n'=1}^n K_2(n'/N)
\end{equation}
where the limit is taken along suitable subsequences $N_1,N_2,\ldots\to\infty$
of integers. To illustrate the relevance of this quantity, let us consider
the counting function $\scrN(L,\xi)$ for the number of eigenphases
in the interval $[\xi-\frac{\pi L}{N},\xi+\frac{\pi L}{N}]$ 
(mod $2\pi$). The number variance is defined by
\begin{align}
\Sigma^2(L) & = \frac{1}{2\pi} \int_0^{2\pi} [\scrN(L,\xi)-L]^2 d\xi \\
& = \frac{L}{N} \sum_{j,k=1}^N \sum_{m\in\ZZ} 
\zeta\left(\frac{N}{2\pi L}(\theta_j-\theta_k+2\pi m)\right) 
-L^2 \label{S2}
\end{align}
with $\zeta(x)=\int_\RR \chi(x-y)\chi(y)dy=\max\{1-|x|,0\}$,
where $\chi$ is the indicator function
of the interval $[-\frac{1}{2},\frac{1}{2}]$.
In view of \eqref{R2}, \eqref{R22} and \eqref{S2}, 
number variance and form factor 
are related by
\begin{equation}
\Sigma^2(L) = \frac{L^2}{N^2} \sum_{n\neq 0} 
\big|\Tr \left[U(\Phi_\alpha)^n\right]\big|^2\;
\widehat\chi\left(\frac{L}{N}\, n\right)^2 
\end{equation}
where $\widehat\chi(y)=\sin(\pi y)/(\pi y)$ 
is the Fourier transform of $\chi$.
It then follows from \eqref{K00} and
a standard probabilistic argument that,
for macroscopic intervals of size $L=\ell N$ (with $\ell>0$ fixed
as $N_\nu\to\infty$), we have 
\begin{equation}\label{LC}
\lim_{\ell\to 0} \lim_{\substack{N_\nu\to\infty\\ L=\ell N_\nu} }
\frac{\Sigma^2(L)}{L}
=
\overline{K_2(0)}
\end{equation}
since
$\ell \sum_{k\neq 0} \widehat\chi(k\ell)^2 \to \int \widehat\chi(y)^2 dy=1$,
as $\ell\to 0$. 
The ratio $\Sigma^2(L)/L$ is called the {\em level compressibility}
\cite{GorWie03}.

\section{Spectral form factor at small argument \label{secSpectral}}

To calculate the value of $K_2(\tau)$ at small but non-zero values of $\tau$,
we note that in the semiclassical limit $N\to\infty$, the trace
$\Tr U(\Phi_\alpha)^n$ is asymptotically equal to $\Tr U(\Phi_\alpha^n)$,
with $n$ arbitrary but fixed. This may be seen by representing the
respective traces as Gutzwiller-type periodic orbit sums. 

The advantage of the choice of $\Phi_\alpha$ over other piecewise
linear maps
is that there is an explicit formula for the $n$th iterate.
(We have observed intermediate statistics also for the 
closely related ``triangle maps''
introduced by Casati and Prosen \cite{CasPro02}; 
the classical analysis of these maps is however considerably more involved.)
Here, a short calculation shows that 
\begin{equation}
\Phi_\alpha^n=\rho_{(n-1)\alpha/2} \circ \Phi_0^n \circ \rho_{(n+1)\alpha/2}.
\end{equation}
The corresponding quantum evolution is therefore given by
\begin{equation}
\label{ufn}
\langle Q'|U(\Phi_\alpha^n)|Q \rangle 
= e\left( N \left(\tfrac{n-1}{2}\right)\alpha \left\{ \tfrac{Q'}{N} \right\} \right) 
\langle Q'|U(\Phi_0^n)|Q \rangle 
e\left( N \left(\tfrac{n+1}{2}\right)\alpha \left\{ \tfrac{Q}{N} \right\} \right) ,
\end{equation}
and so
\begin{equation}
\Tr U(\Phi_\alpha^n) = \frac{1}{N} \sum_{P=0}^{N-1} e_N(-n P^2)
\times \sum_{Q=0}^{N-1} e( n \alpha Q)
\end{equation}
where the first sum is a classical Gauss sum and the
second a geometric sum. Let $m:= 2n / \gcd(2n,N)$
and $M:= N / \gcd(2n,N)$, then
\begin{equation}
\left| \sum_{P=0}^{N-1} e_N(-n P^2) \right| = 
\frac{N}{M}
\left| \sum_{P=0}^{M-1} \exp\left(\pi\i P^2 \tfrac{m}{M}\right) \right|
\end{equation}
which evaluates to $N/\sqrt M$ if $Mm$ is divisible by 2, and 
vanishes otherwise. 
The geometric sum is $O(1)$ for irrational $\alpha$ and $n\neq 0$
and hence $K_2(n/N)\sim 0$ for all bounded $n$ in this case. 
For rational $\alpha=a/b$, the geometric sum equals $N$ if $n$
is divisible by $b$ and is $O(1)$ otherwise.
Thus $K_2(n/N)\sim \gcd(2n,N)$ provided $n$
is divisible by $b$ and $2Nn/\gcd(2n,N)^2$ is divisible by 2;
$K_2(n/N)\sim 0$ in all other cases.

If we restrict ourselves to a subsequence of the values of $N$
which are prime numbers then $\gcd(2n,N)=1$ for $n<N$, 
and the time averaged form factor is
\begin{equation}\label{K0}
\overline{K_2(0)}
:=
\lim_{n\to\infty} \lim_{\substack{N\to\infty\\ \text{$N$ prime}}} 
\frac1n \sum_{n'=1}^n K_2(n'/N)
= \frac1b .
\end{equation}
These values of $\overline{K_2(0)}$ are consistent with those expected for 
intermediate statistics \cite{BogGirSch01}.
The case $b=2$ and $N$ prime, for which $\overline{K_2(0)}=1/2$, does 
however not agree with the Poisson statistics seen numerically in
Fig.~\ref{Ps1} (left), where  $\overline{K_2(0)}=1$.
The solution to this apparent paradox is that $P(s)$ 
displays correlations on the scale
of the mean level spacing, whereas $\overline{K_2(0)}$ involves
correlations on much larger scales of order $N$. Similar discrepancies
between a random matrix-like $P(s)$ 
and a non-universal $K_2(\tau\approx 0)$  have been 
observed for non-arithmetic Hecke triangles \cite{BogGeoGiaSch97,BogSch04},
compact hyperbolic triangles and tetrahedra \cite{AurMar03}
and cat maps coupled to a spin $1/2$ \cite{KepMarMez01}.

If $N$ is twice a prime, i.e., $N=2 R$ with $R$ an odd prime, 
then $\gcd(2n,N)=2$ for $n<R$, and $2Nn/4=Rn$ is divisible
by 2 if and only if $n$ is even. Hence only terms with $n$ divisible
by $\lcm(2,b)$ contribute, and so 
\begin{equation}\label{K0a}
\overline{K_2(0)}
:=
\lim_{n\to\infty} \lim_{\substack{N\to\infty\\ \text{$N/2$ prime}}}
\frac1n \sum_{n'=1}^n K_2(n'/N)
= \frac{2}{\lcm(2,b)} .
\end{equation}

The situation is different for $N=\rho R$, where $\rho$ is a fixed odd prime,
and $R$ runs again over all odd primes. Now $\gcd(2n,N)=\gcd(n,\rho)$ for
$n<R$, and $2Nn/\gcd(n,\rho)^2$ is always divisible by 2.
Since $\rho$ is prime, $\gcd(n,\rho)=1$ if $n$ is not divisible by $\rho$
and $\gcd(n,\rho)=\rho$ if it is. In this case
\begin{equation}\label{K0b}
\overline{K_2(0)}
:=
\lim_{n\to\infty} \lim_{\substack{N\to\infty\\ \text{$N/\rho$ prime}}} 
\frac1n \sum_{n'=1}^n K_2(n'/N)
= \frac1b + \frac{\rho-1}{\lcm(\rho,b)} .
\end{equation}

\begin{figure}
\label{compressibility}
\begin{center}
\includegraphics[width=0.30\textwidth,angle=-90]{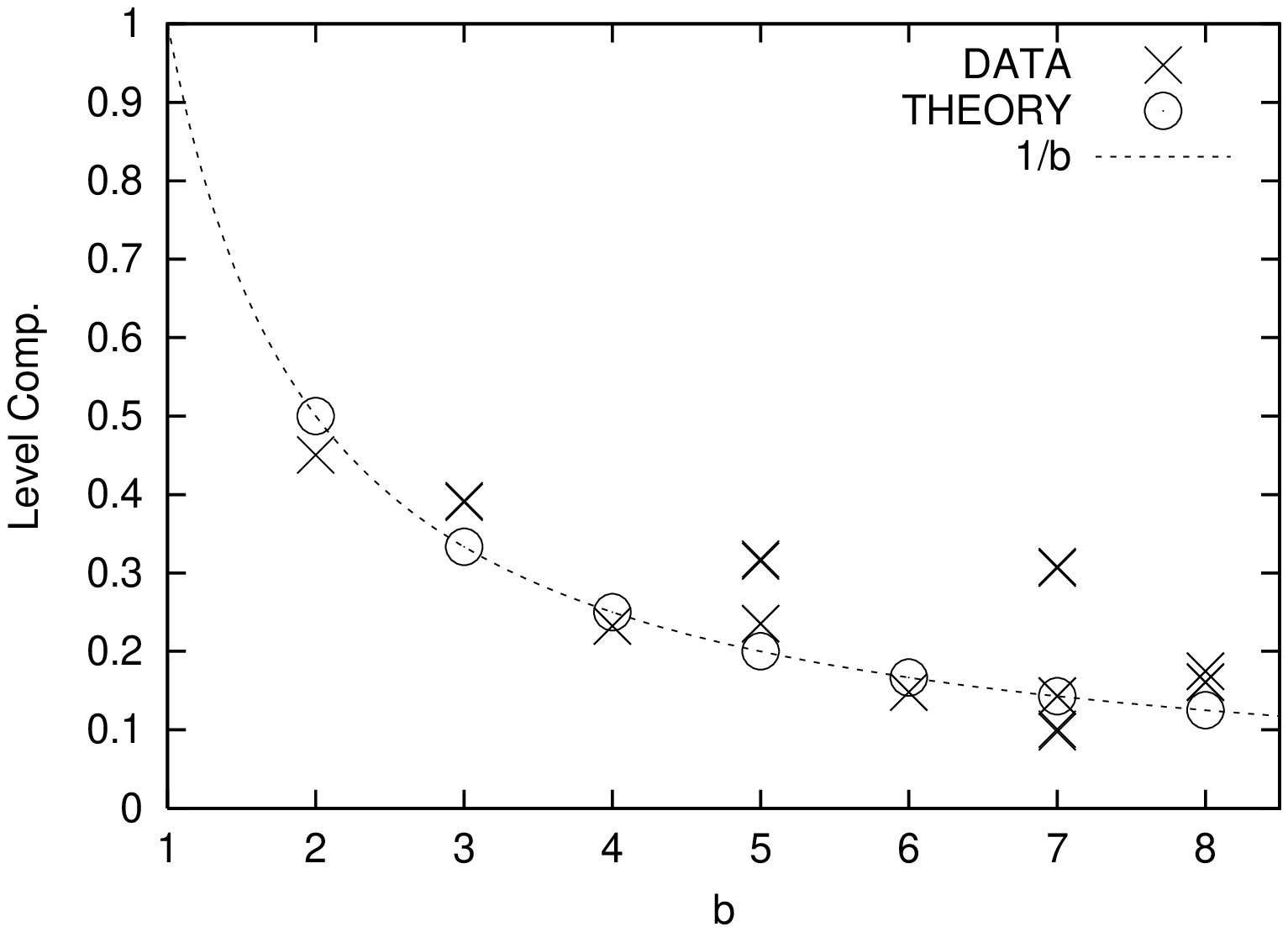}
\includegraphics[width=0.30\textwidth,angle=-90]{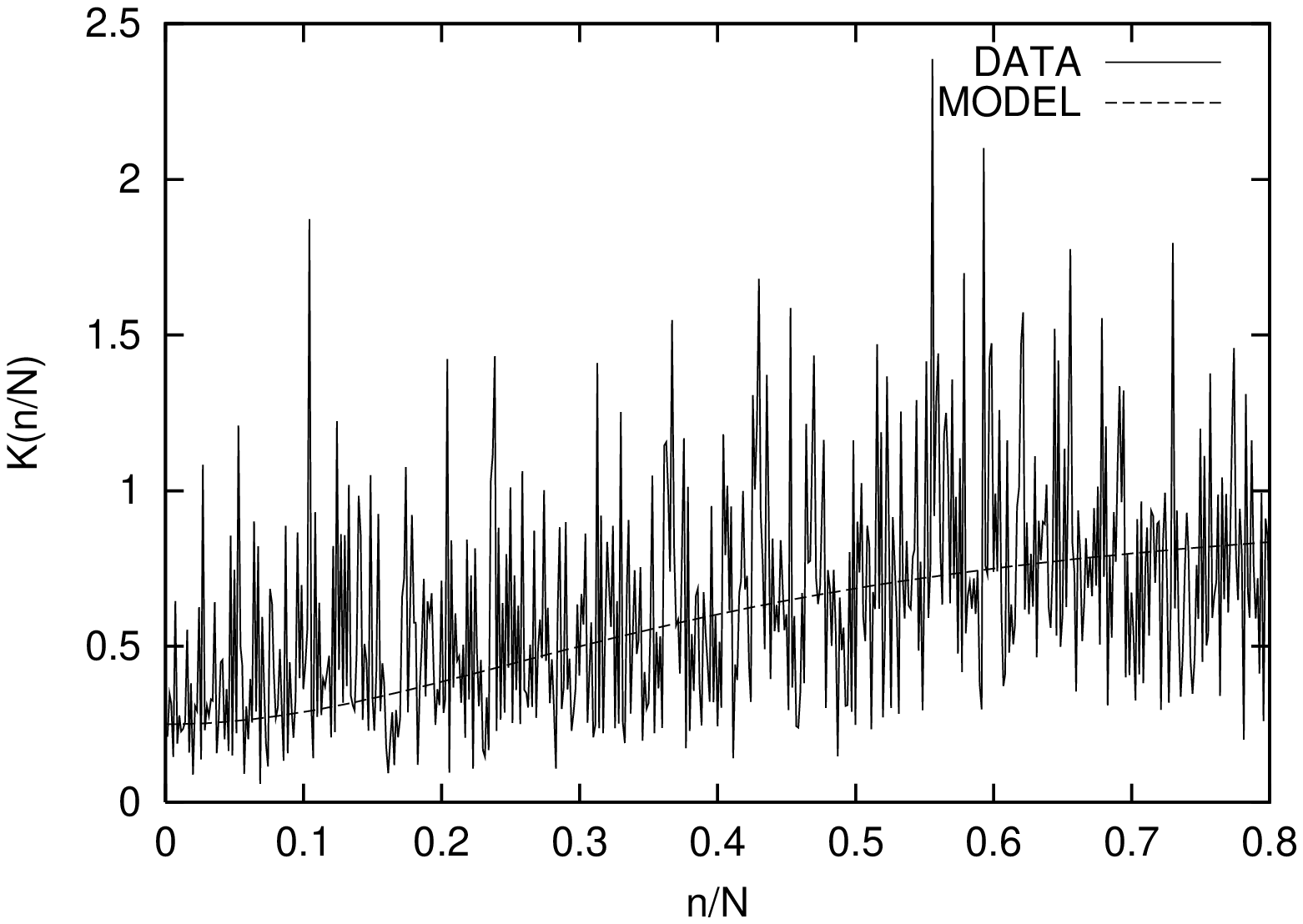}
\includegraphics[width=0.30\textwidth,angle=-90]{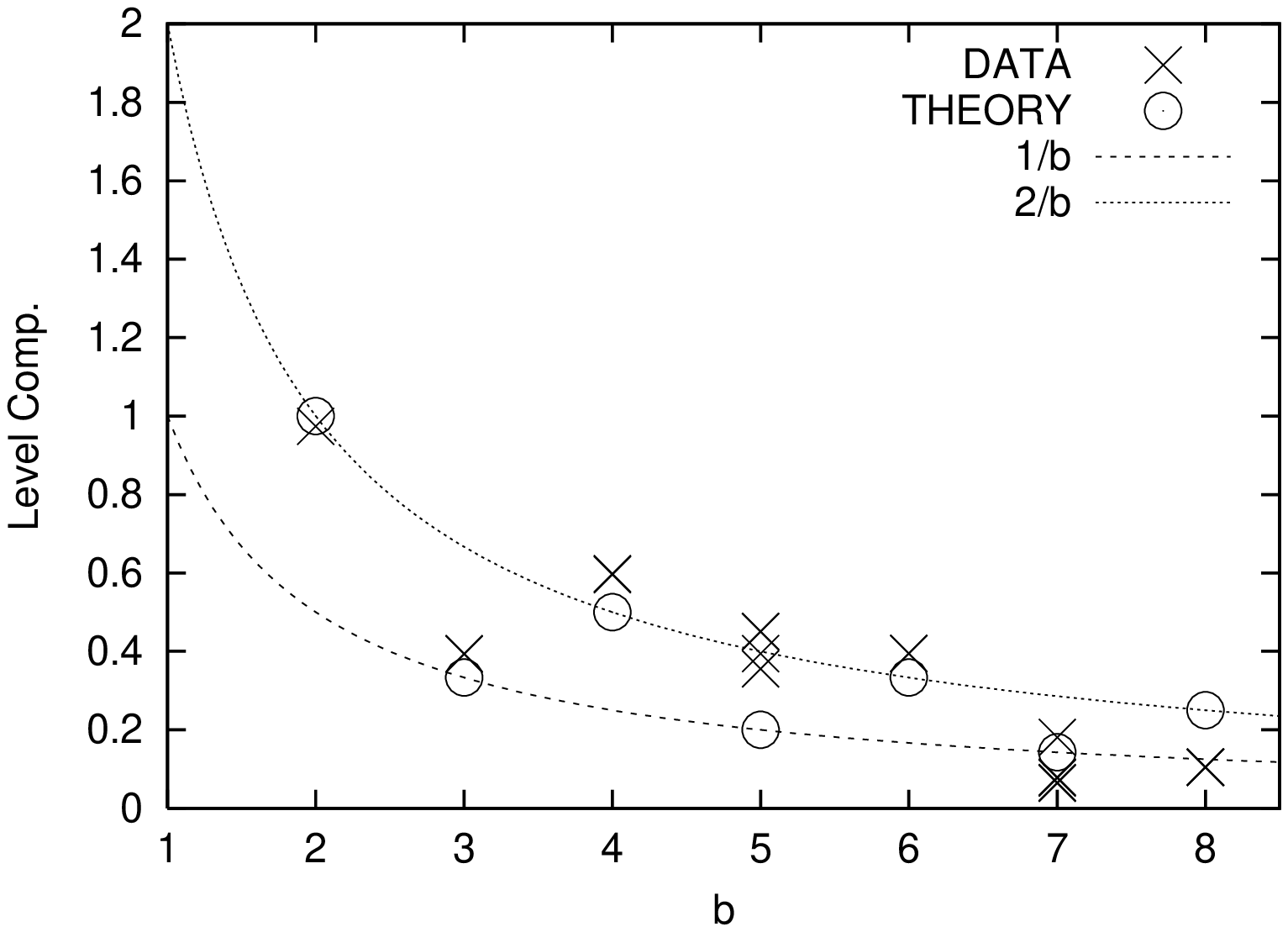}
\includegraphics[width=0.30\textwidth,angle=-90]{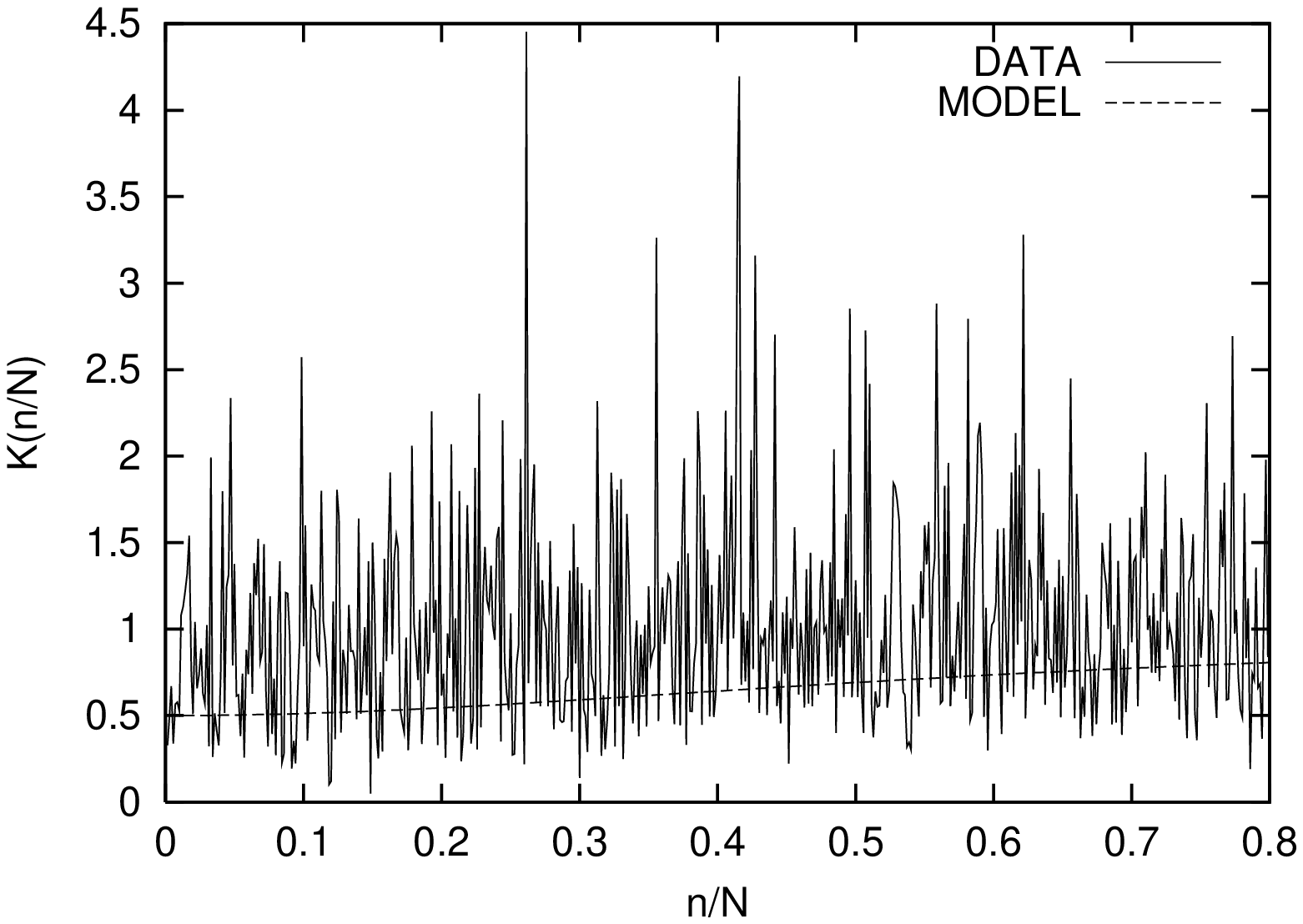}
\includegraphics[width=0.30\textwidth,angle=-90]{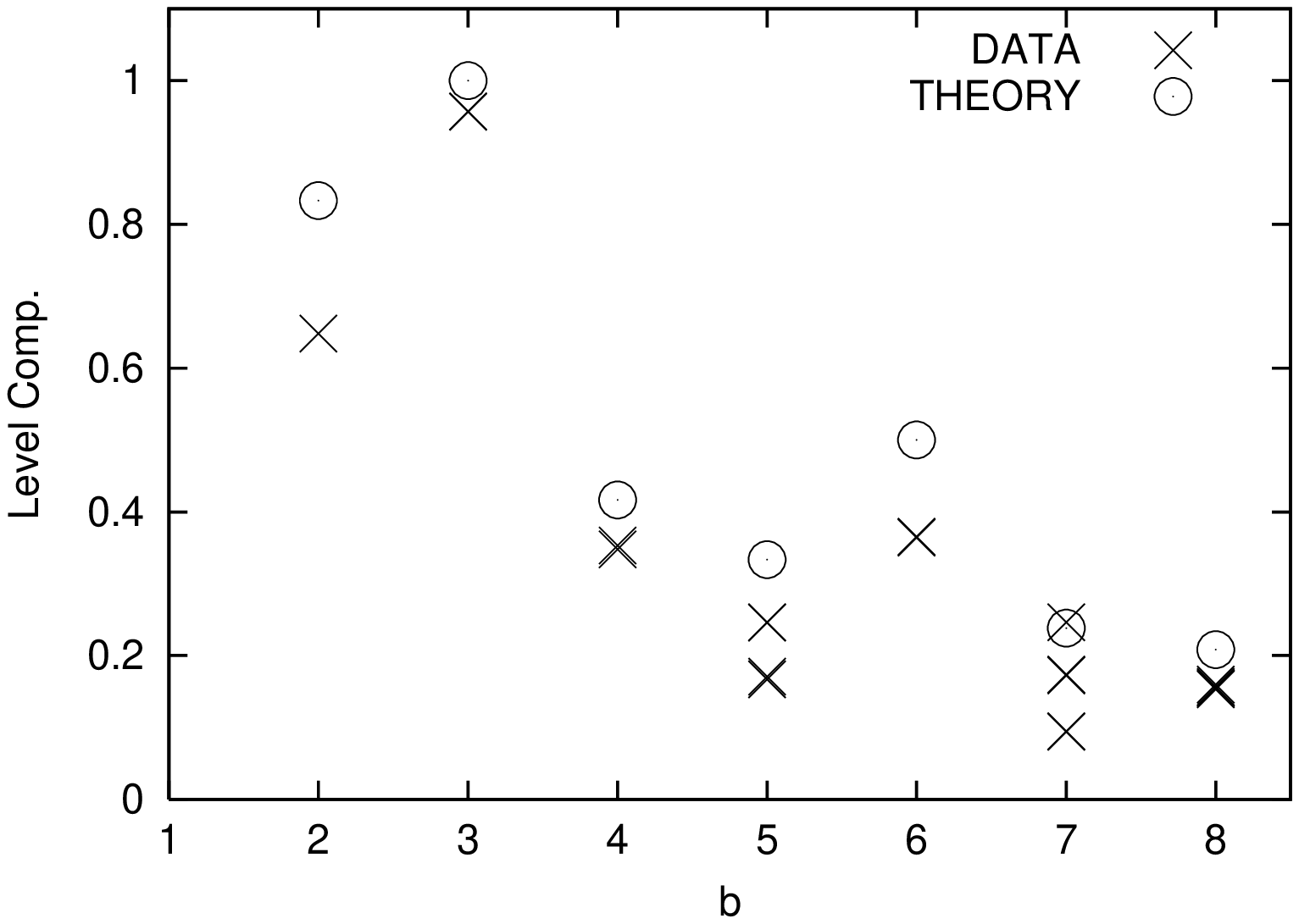}
\includegraphics[width=0.30\textwidth,angle=-90]{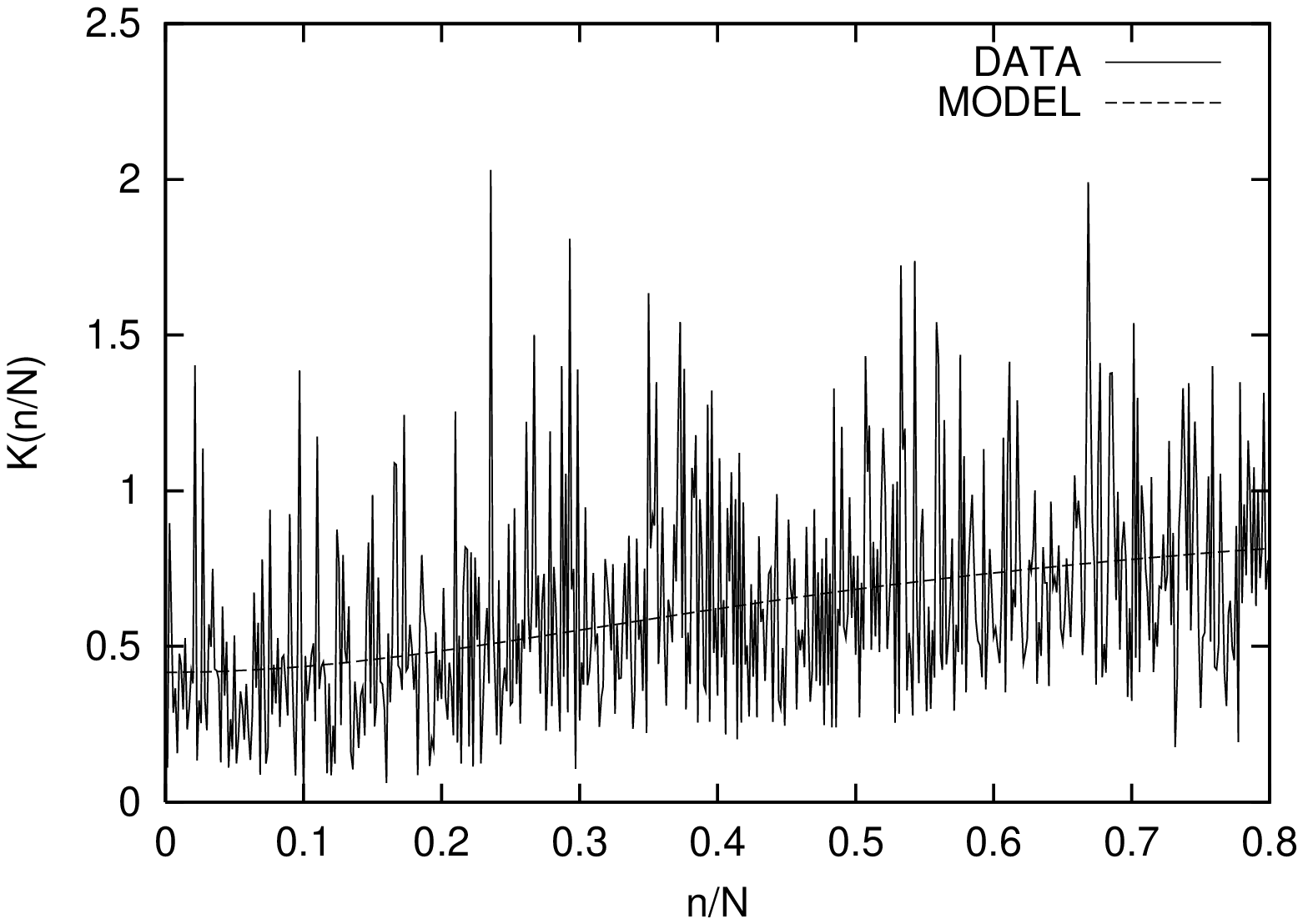}
\caption{Left: 
Level compressibility $\Sigma^2(L)/L$ with $L=100$
for rational $\alpha=a/b$ ($\times$), 
versus $\overline{K_2(0)}$ ($\odot$). 
From top to bottom: $N=7001$
(Equation (\ref{K0})), $N=6998$ (Equation (\ref{K0a})), and $N=6999$ (Equation (\ref{K0b})).
Right: numerical value of the form factor and theoretical plot (\ref{formfactorsp}) with
$K_2(0)$ given by  $\overline{K_2(0)}$, for the same $N$ and $\alpha=1/4$.}
\end{center}
\end{figure}

Fig.~\ref{compressibility} (left) illustrates the asymptotic relation
\eqref{LC} between the level compressibility and
$\overline{K_2(0)}$ for large values of
$N$: a prime ($N=7001$), twice a prime ($N=6998$) 
and three times a prime ($N=6999$).
Fig.~\ref{compressibility} (right) compares the numerical form factor,
obtained by diagonalizing the matrices $U(\Phi_\alpha)$,
with the model \cite{BogGirSch01}
\begin{equation}
\label{formfactorsp}
K_{2}(\tau)=\frac{\lambda^2-2\lambda+4\pi^2 \tau^2}{\lambda^2+4\pi^2 \tau^2}
\end{equation}
with $\lambda$ equal to $2/(1-\overline{K_2(0)})$.
If $N$ is divisible by $b$, the operator
$U(\Phi_\alpha)$ coincides with an alternative quantization
$U_0(\Phi_\alpha):=U(\Phi_{A/N})$
of $\Phi_\alpha$ proposed in \cite{MarRud00}, 
where $\alpha$ is replaced by a rational approximation
$A/N$ so that $|\alpha-A/N|<1/N$.
The spectral statistics of $U_0(\Phi_\alpha)$ are 
well known to be highly singular and are not of intermediate type
\cite{BacHaa99}. It has been noted in \cite{HaaKep02, Kep03}
that $U_0(\Phi_\alpha)$ may be coupled to a spin $1/2$ precession
in such a way that intermediate statistics are seen numerically; the
construction is analogous to the one for 
cat maps with spin $1/2$ \cite{KepMarMez01}.

\section{Localized level spacing distributions \label{secNon}}

The above analysis yields trivially for irrational $\alpha$
\begin{equation}
\overline{K_2(0)}
:=
\lim_{n\to\infty} \lim_{\substack{N\to\infty}} 
\frac1n \sum_{n'=1}^n K_2(n'/N)
= 0
\end{equation}
consistent with the COE statistics seen in Fig.~\ref{Ps3} (left). 
In contrast, Fig.~\ref{Ps3} (right) illustrates a class of
statistics different from random matrix theory, which occur
for subsequences of $N$, for which the quantity 
\begin{equation}
\epsilon:= 
\begin{cases}
\{N\alpha\} & \text{if $\{N\alpha\} \leq 1/2$}\\
\{N\alpha\}-1  & \text{otherwise}
\end{cases} 
\end{equation}
(the oriented distance of $N\alpha$ to the nearest integer)
is at most of the order of $1/\sqrt N$. Note that if we take 
$A/N$ to be the successive approximants in the continued fraction expansion
of $\alpha$ irrational, we have $\epsilon=O(1/N)$. 
On the other hand, for rational
$\alpha=a/b$ with $N$ not divisible by $b$, we
have $|\epsilon|\geq 1/b$; the following considerations
clearly to not apply in the latter case, where we may expect to
see generic intermediate statistics.

\begin{figure}
\begin{center}
\includegraphics[width=0.37\textwidth,angle=-90]{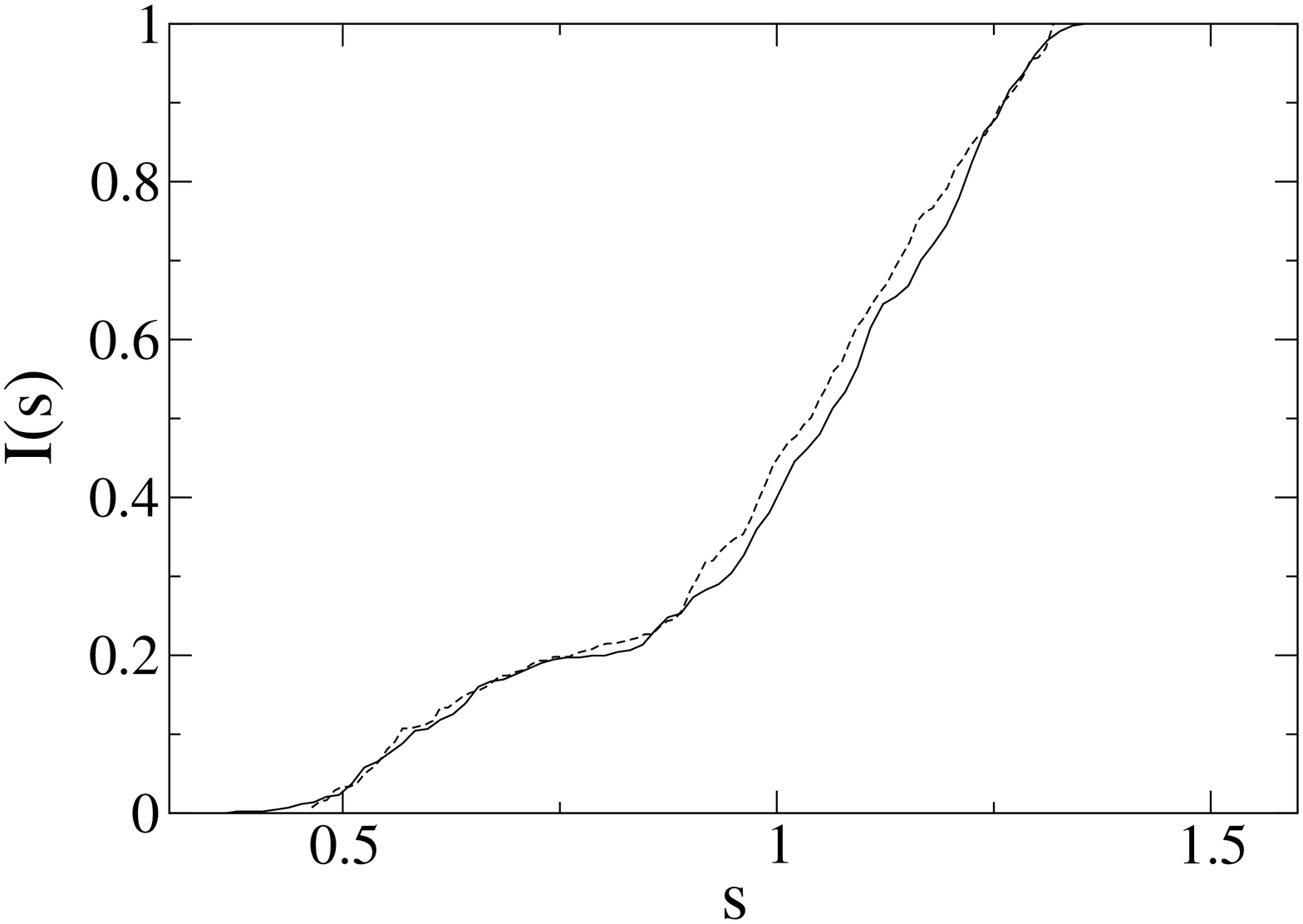}
\includegraphics[width=0.37\textwidth,angle=-90]{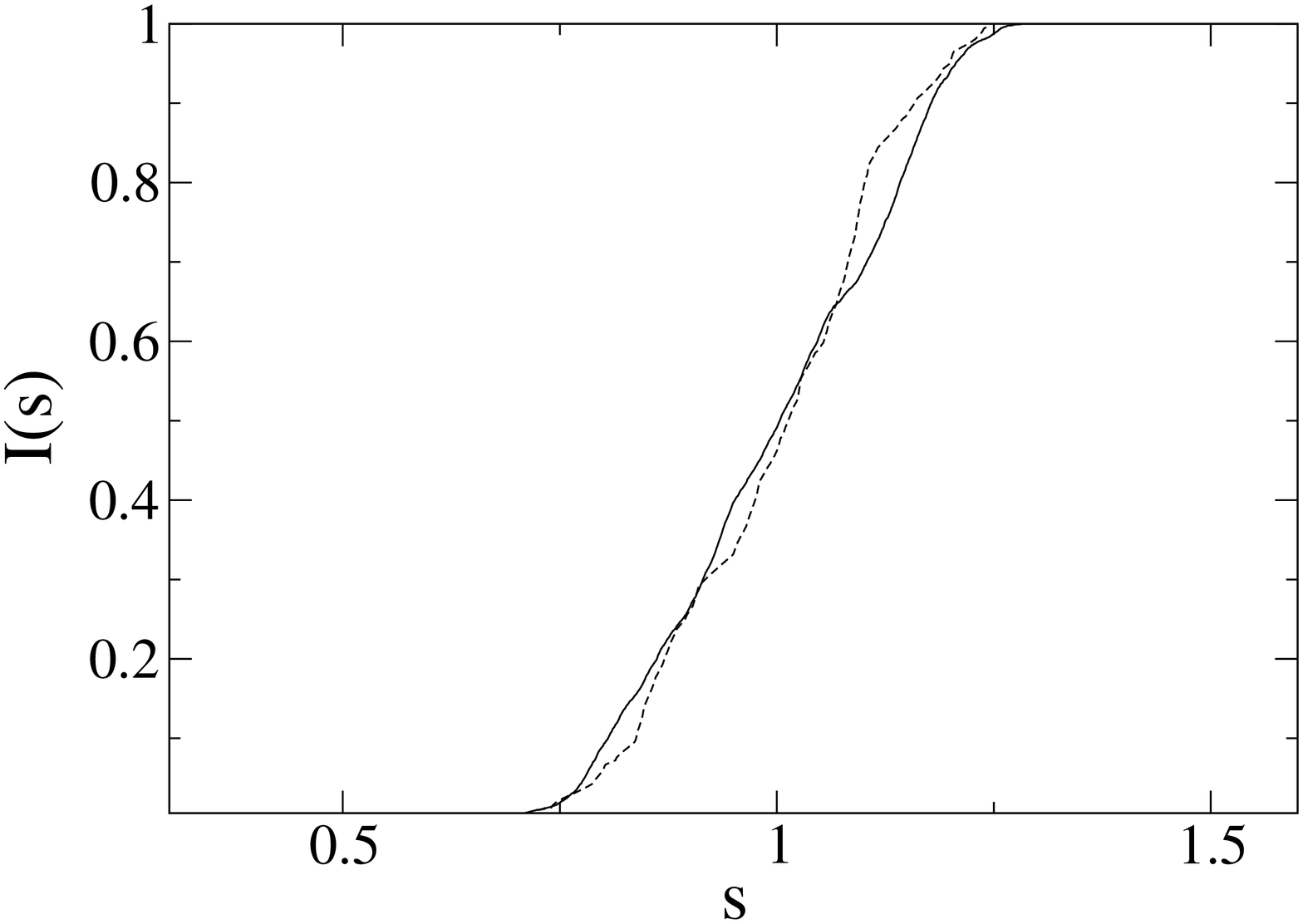}
\caption{Integrated level spacing distribution for $U(\phi)$ (solid line) and integrated distribution
 of $G(\phi)$ for $\phi$ uniformly distributed random variable (dashed line) for $N=431$ (left) and
$N=5867$ (right).}
\label{bornrandom}
\end{center}
\end{figure}

We shall now explain the localized level correlations observed for
$\epsilon=O(1/\sqrt N)$ by means of classical perturbation theory.
The eigenphases $\theta_1^{(0)},\ldots,
\theta_N^{(0)}\in[0,2\pi)$ of $U_0(\Phi_\alpha)$ 
and the corresponding orthonormal basis of eigenstates
$\varphi_1^{(0)},\ldots,\varphi_N^{(0)}$
are known explicitly, cf.~\cite{MarRud00}, Prop.~5.1. Since
\begin{equation}
\langle Q'|U(\Phi_\alpha)|Q \rangle =\langle Q'|U_0(\Phi_\alpha)|Q \rangle \;
e_N(\epsilon Q),
\end{equation}
the Born expansion of the eigenphases $\theta_j$ of $U(\Phi_\alpha)$ is
$\theta_j=\theta_j^{(0)}+\epsilon\theta_j^{(1)}+O(\epsilon^2)$
with the first order correction given by
\begin{equation}
\label{1storder2}
\theta_j^{(1)}=\pi + 2\pi
\langle \varphi_j^{(0)} | \Delta | \varphi_j^{(0)} \rangle,
\end{equation}
where $\Delta(\phi)=\{\phi\}-1/2$ is the sawtooth function.
The term $\pi$ is irrelevant for the spacing distribution since 
it is independent of $j$. As to the second term,
quantum unique ergodicity of $U_{0}(\Phi_\alpha)$, proved in \cite{MarRud00},
implies that in the limit $N\to\infty$ we have
$\langle \varphi_j^{(0)} | \Delta | \varphi_j^{(0)} \rangle
\to \int_0^1 \Delta(\phi)\, d\phi = 0$ for all $j$.
The eigenstates $\varphi_j^{(0)}$ have a particularly simple form
for $N$ a prime number \cite{MarRud00}, which we will assume in the following.
In this case it can be shown that the level spacing distribution $P(s)$ for the $\theta_j$ is
asymptotically given by the distribution of
\begin{equation}
G(\phi)=1+\epsilon \sqrt N\sum_{\substack{k\in\ZZ \\ k\neq 0}}
\frac{1-e_N(-A^{-1}k)}{2\pi i k}
\,e(k\phi+\beta_{k,N})
\end{equation}
where $\phi$ is a uniformly distributed random variable in
$[0,1)$, $A^{-1}$ the inverse of $A$ modulo $N$ and $\beta_{k,N}$ some 
explicitly known phase factor (see Appendix). The variance of the above 
distribution is $\epsilon^2 A^{-1}(1-A^{-1}/N)$, with the choice 
$A^{-1}\in[0,N-1]$.
Fig.~\ref{bornrandom} compares the numerical computation of the 
integrated level-spacing
distribution $I(s)=\int_0^s P(s')ds'$
for the eigenvalues $\theta_j$ of $U(\Phi_\alpha)$ 
(solid line) to 
the distribution of the random variable $G(\phi)$ (dashed line).

\section*{Acknowledgments}
We thank E. Bogomolny for stimulating discussions.
This research is supported by 
EPSRC Research Grant GR/R67279/01 (J.M. \& S.O'K.),
an EPSRC Advanced Research Fellowship (J.M.),
the Leverhulme Trust (O.G.), and the
EC Research Training Network (Mathematical Aspects of Quantum Chaos) 
HPRN-CT-2000-00103. The numerical computations have
been performed on an Alpha Workstation funded by a Royal Society
Research Grant.

\section*{Appendix}
Let us define $\beta_{k,N}$ by
\begin{equation}
e(\beta_{k,N})=e_N\left(A^2\sum_{r=1}^{A^{-1}k}r^2-\frac{Ak}{N}\sum_{r=1}^{N}r^2\right)
\frac{1}{\sqrt{N}}\sum_{\nu=0}^{N-1}e_N\left(A k \nu^2+(A+k)k\nu\right).
\end{equation}
It can be shown that the variable
$\xi_j:= \sqrt N \langle \varphi_j^{(0)} | \Delta | \varphi_j^{(0)} \rangle$
is equal to $\xi_j=h(A^{-1}j/N)$, where $h$ is the 1-periodic function
\begin{equation}
h(x)=\sum_{k\in\ZZ} \widehat\Delta_{-k}e(k x+\beta_{k,N}),
\end{equation}
where $\widehat\Delta_0=0$,
$\widehat\Delta_k=i/2\pi k$ ($k\neq 0$)
are the Fourier coefficients of $\Delta(\phi)$.
For $N$ prime, the spectrum of $U_0(\Phi_\alpha)$
is totally rigid \cite{MarRud00}, that is
$\theta^{(0)}_j=2\pi j/N+C_N$ for $j=1,\ldots,N$, where $C_N$ is some overall shift.
The corresponding level spacing distribution is hence 
$P^{(0)}(s)=\delta(s-1)$. For $\epsilon$ small enough, the perturbation does
not change the ordering of the levels. The  level spacing distribution $P(s)$ for the $\theta_j$ is
then given by the distribution of the 
$(N/2\pi)(\theta_j-\theta_{j-1})\approx 1+ \epsilon \sqrt N (\xi_j -\xi_{j-1})$,
which is equal to $G(A^{-1}j/N)$, where $G$ is the function defined by
\begin{equation}
G(\phi)=1+\epsilon \sqrt N\sum_{k\in\ZZ} \widehat\Delta_{-k}\,e(\beta_{k,N})
\left[1-e_N(-A^{-1}k)\right]\,e(k\phi).
\end{equation}
For random $j\in\{1,\ldots,N\}$ and $N$ large the distribution of $P(s)$ is 
asymptotically given by the distribution of $G(\phi)$
where $\phi$ is now a uniformly distributed random variable in
$[0,1)$. The variance of the above distribution is 
$(2\epsilon\sqrt N)^2\sum_{k\in\ZZ} 
|\widehat\Delta_k|^2\sin^2\left(\pi A^{-1}k/N\right)=
\epsilon^2 A^{-1}(1-A^{-1}/N)$, provided we choose $A^{-1}\in[0,N-1]$.

\end{document}